\documentstyle[12pt,epsf]{article}
\newcommand{\gsim}{ \raisebox{-.5ex}{\mbox{$\,\stackrel{>}{\sim}\,$}} }
\newcommand{\lsim}{ \raisebox{-.5ex}{\mbox{$\,\stackrel{<}{\sim}$\,}} }
\begin{document}
\begin{center}
{\large\bf Few--Freedom Turbulence}\\[2mm]
B.V. Chirikov\footnote{Email: chirikov @ inp.nsk.su} and V.G. Davidovsky\\[2mm]
{\it Budker Institute of Nuclear Physics \\
        630090 Novosibirsk, Russia}\\[5mm]
\end{center} 
\baselineskip=15pt

\vspace{2mm}

\begin{abstract} 
The results of numerical experiments on the structure of chaotic attractors
in the Khalatnikov - Kroyter model of two freedoms 
are presented.
This model was developed for a qualitative description
of the wave turbulence of the second sound in helium.
The attractor dimension, size, and the maximal Lyapunov exponent
in dependence on the single dimensionless
parameter $F$ of the model are found and discussed.
The principal parameter $F$ is similar to the Reynolds number
in hydrodynamic turbulence.
We were able to discern four different attractors characterized by
a specific critical value of the parameter ($F=F_{cr}$), such that the 
attractor exists for $F>F_{cr}$ only.
A simple empirical relation for this dependence on the 
argument ($F-F_{cr}$) is presented which turns out
to be universal for different attractors with respect to the dimension
and dimensionless Lyapunov exponents.
Yet, it differs as to the size of attractor.
In the main region of our studies the dependence of all dimensionless 
characteristics of
the chaotic attractor on parameter $F$ is very slow (logarithmic) which
is qualitatively different as compared to that of a 
multi--freedom
attractor, e.g., in hydrodynamic turbulence (a power law).
However, at very large $F\sim 10^7$
the transition to a power--law dependence has been finally found,
similar to the multi--freedom attractor.
Some unsolved problems and open questions are also discussed.
\end{abstract}

\section{Introduction}
In the present paper we continue numerical experiments \cite{1} with the
Khalatnikov - Kroyter model \cite{2} developed for a qualitative description
of the wave turbulence of second sound in helium.

The model is specified by an
effective non--Hermitian Hamiltonian:
$$
   H(a_1,\,a_2)\,=\,(\omega_1\,-\,i\gamma_1)|a_1|^2\,+\,
   (\omega_2\,-\,i\gamma_2)|a_2|^2\,
   +\,\left(\mu a_1^2a_2^*\,+\,fa_1^*\,
   +\,{\rm c.c.}\right) \eqno (1.1)
$$
where we slightly changed the notations in Ref.\cite{2,1}.
This Hamiltonian describes the two linear oscillators via complex 
phase--space variables $a_j\ (j=1,2)$ and the 
frequencies $\omega_j-i\gamma_j$
with phenomenological dissipation parameters $\gamma_j$.
Two other parameters of the model represent a 
nonlinear coupling $\mu$ of the two oscillators, and the driving
force $f$.

The motion equations 
$$
\begin{array}{lll}
 i \dot a_1 & = & (\omega_1-i \gamma_1) a_1+2\mu a_1^* a_2+f \\
 i \dot a_2 & = & (\omega_2-i \gamma_2) a_2+\mu a_1^2
\end{array}
\eqno (1.2)
$$
were numerically integrated together with the corresponding linearized
equations (for details see Ref.\cite{1}).
Particularly, all four Lyapunov exponents
($\lambda_1\geq\lambda_2\geq\lambda_3\geq\lambda_4$) were computed.
One of them, whose eigenvector goes along the 
trajectory, is always zero while the sum of all 
$$
   \lambda_{\Gamma}\,=\,\sum_{n=1}^4\,\lambda_n\,=\,
   -2(\gamma_1\,+\,\gamma_2)\,=\,
   {\rm const} \eqno (1.3)
$$ 
is the constant rate of the phase space volume contraction.

Surprisingly, at a relatively weak force $f$ this most simple model does
describe the birth of turbulence (dynamical chaos) in a physical system.
This was the main subject of studies in Ref.\cite{2,1}.

As the force grows the two--freedom model is losing any
relation to the real physical turbulence which is 
a multi(infinitely)--freedom
phenomenon. Nevertheless, it seems of some interest, 
for the general theory
of dynamical systems, to compare the structure and properties of a chaotic
attractor in such an opposite limit represented by a sample model under
consideration. It was the main motivation for us to continue general studies
of the model on unrestricted range of its parameters.

\section{Scaling}
In the spirit of the theory of turbulence we introduce, first,
a unique dimensionless parameter, similar to the Reynolds number,
which determines all the dimensionless characteristics
of the motion. 
To this end, we choose $f,\ \mu ,\ \gamma$ as the three basic parameters,
and form the desired combination 
$$
   F\,=\,\frac{f\mu}{\gamma^2} \eqno (2.1)
$$
which has the meaning of a dimensionless force. For this basic characteristic
of the model were unique we need to fix all other dimensionless parameters:
$$
  \frac{\omega_1}{\omega_2}\,=\,0, \quad 
  \frac{\gamma_1}{\gamma_2}\,=\,0.25, \quad {\rm and} \quad
  \frac{\omega_2}{\gamma_2}\,=\,-12.5 \eqno (2.2)
$$
The values of all three ratios are chosen as in the previous studies 
\cite{2,1}. Particularly, we keep $\omega_1=0$, as well as the motion initial 
conditions
$$
   a_1(0)\,=\,a_2(0)\,=\,0 \eqno (2.3)
$$
This allows us to directly compare our new results with the former ones.
In what follows
we set $\gamma =\gamma_2$ in the basic relation (2.1).

Invariance of $F$ with respect to variation of the three basic
parameters gives us an extra freedom for choosing the latter
in such a way to minimize the computation errors, and thus to
reach higher values of $F$. 

In the present studies we were primarily interested in the properties
of chaotic attractor. One of its principal characteristic is
the metric ($M$) that is the set of signs of the Lyapunov exponents.
As in the previous studies we did observe only one metric of the three
possible, namely
$$
   M\,=\,(+,\,0,\,-,\,-) \eqno (2.4)
$$
with a single zero exponent $\lambda_2=0$.
The next, more interesting, characteristic of attractor is the fractal
(noninteger) dimension. By now, there is a dozen of various definitions
for such a dimension (see, e.g., Ref.\cite{3}).
From physical point of view those all are meaningful and
acceptable in principle.
However, in numerical experiments the two of them are much more preferable.
These make use of the Lyapunov exponents only which greatly simplifies the
computation. We have chosen one, due to Kaplan and Yorke, because it is
more close to other definitions \cite{4,5}.
It is given by the relation
$$
   d\,=\,m\,+\,\frac{\sum_{n=1}^m\,\lambda_n}{|\lambda_{m+1}|}
   \eqno (2.5a)
$$
where $m$ is the largest integer for which the sum $\sum_{n=1}^m \lambda_n
\geq 0$. This is the simplest and widely used method for calculating
dimension.
In the model under consideration
the dimension of a chaotic attractor is always within the interval
$(2<d<4)$. The upper bound corresponds to the dimension of the whole
phase space, while the lower is the condition for chaos ($\lambda_1>0$,
see Eq.(2.4)).

The second characteristic of a chaotic attractor, closely related to
(but different from) the former is the maximal Lyapunov exponent
$\lambda_1$. In dimensionless form it is
$$
   \Lambda_1\,=\,\frac{\lambda_1}{\gamma} \eqno (2.5b)
$$
Accordingly, the dimensionless sum of all the exponents $\Lambda_{\Gamma}
=\lambda_{\Gamma}/\gamma =-2.5$ (see Eq.(1.3)). 

Besides, we computed two other, geometrical, characteristics of attractor:
the average size and shape. The former is represented by attractor's 
rms radius
in phase space
$$
   R\,=\,\sqrt{R_1^2\,+\,R_2^2}\,, \quad
   R_j^2\,=\,<A_j^2>\,-\,<A_j>^2\,\approx\,<A_j^2> \eqno (2.5c)
$$
where $A_j=a_j\sqrt{\mu /f}\ (j=1,2)$ are dimensionless variables, and the
brackets $<>$ denote the time average along a trajectory.
In all cases the average shift of attractors $<A_j>$ was
relatively small.
The attractor shape is characterized by the ratio
$$
   S\,=\,\frac{R_2}{R_1} \eqno (2.5d)
$$

The results are presented and discussed in the next Section.

\section{Results and discussion}
In Fig.1 the attractor dimension $d$, Eq.(2.5a), is plotted in dependence
on the principal parameter $F$, Eq.(2.1). The huge range of $F$ comprises
almost five orders of magnitude, from the first chaos border at 
$F=F_{cr}=173$ up to $F\approx 10^7$ limitted by the computation time
(about one full day).
Five separated groups of points
are clearly seen followed by a number of scattered points.
Each group corresponds to a chaos window in $F$ separated from the neighbours
by the limit--cycle windows. Apparently, the latter also exist in the region
of scattered points which all belong to chaotic attractors. However, these
points are too few, because of computation difficulties, for the 
reliable location
of any windows beyond $F\sim 10^4$. 

\begin{figure}[h]
\centerline{\epsfxsize=15cm \epsfbox{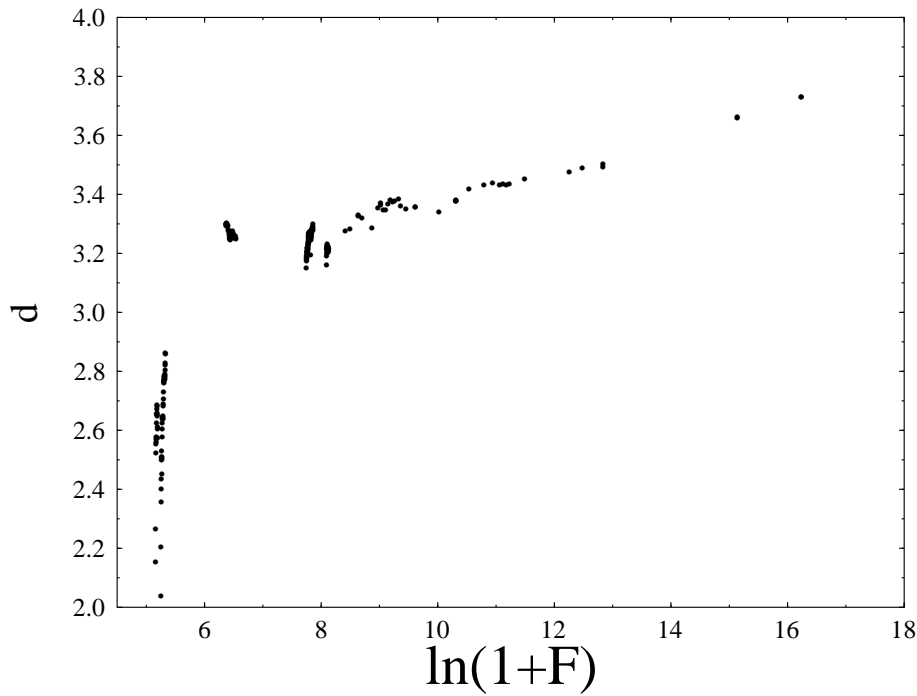}}
\caption{Chaotic attractor dimension vs. principal parameter $F$
         in a semilog scale (see text). Gaps between groups of points
         correspond to limit--cycle attractors.
}
\end{figure}

The first two groups, studied already in our previous work \cite{1},
certainly belong to different attractors with different chaos borders
$F_{cr}\approx 173\ {\rm and}\ 190$, respectively.
It is natural, again in the spirit of the theory of turbulence, to introduce
the principal argument of all dependences in the form:
$$
   \delta F\,=\,F\,-\,F_{cr} \eqno (3.1)
$$
where $F_{cr}$ is the corresponding chaos border which is characteristic, 
and different,
for a particular attractor.

\begin{figure}[h]
\centerline{\epsfxsize=15cm \epsfbox{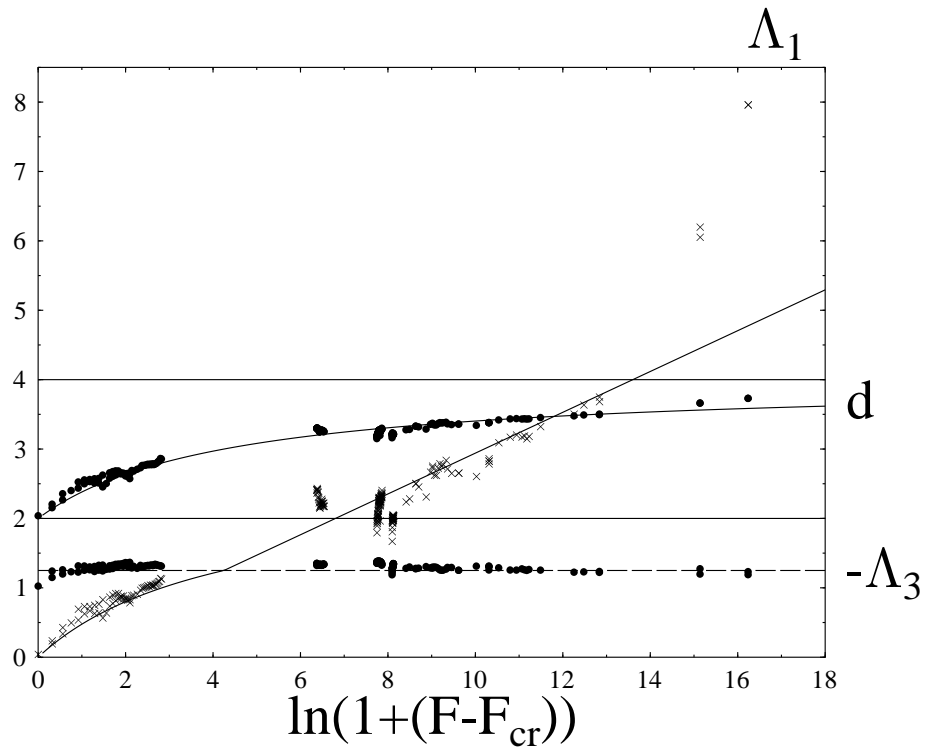}}
\caption{Dependence of the attractor dimension $d$ (upper points),
         maximal Lyapunov exponent $\Lambda_1$ (crosses), and
         $-\Lambda_3$ (lower points) on the parameter $\delta F$ (3.1).
         In this figure $F_{cr}=173\ {\rm and}\ 190$ for the two first groups
         of points in Fig.1, and $F_{cr}=0$ for all other points
         (cf. Figs.1 and 3). Two solid horizontal lines indicate
         the dimension interval ($2<d<4$). The dashed line is
         the mean $<-\Lambda_3>\approx 1.25$.
         Two curves are empirical relations (3.3) and (3.2) 
         for $d$ and $\Lambda_1$, respectively; $F_0=8.5$.
}
\end{figure}

The result of such a rescaling is shown in Fig.2: both groups have overlapped
following some joint dependence $d(\delta F)$.

The first surprise was in that one of the Lyapunov exponents
proved to be nearly constant in the whole range of the empirical data:
$\Lambda_3\approx -1.25\approx\Lambda_{\Gamma}/2$. Even though 
the origin of this peculiarity 
is not known it allows for a simple relation of both empirical dependences in
Fig.2. Besides, we did use this result as an additional check 
of the computation accuracy.

Indeed, using Eqs.(1.3), (2.4), and (2.5a) we obtain
a simple relation
$$
   \Lambda_1\,=\,\left\{
\begin{array}{ll}
   -\Lambda_3\,(d\,-\,2)\,, & d\,\leq\, 3 \\
   -\Lambda_3\,-\Lambda_{\Gamma}\frac{d\,-\,3}{4\,-\,d}\,, &
   d\,\,\geq\,3
\end{array}
   \right .
\eqno (3.2)
$$
with constant parameters $\Lambda_3\approx -1.25$ and 
$\Lambda_{\Gamma}=-2.5\approx 2\Lambda_3$
whatever the dependence $d(\delta F)$.

The next, and the main, result was a very slow dependence on $F$
of all the dimensionless characteristics of chaotic attractors
(see Figs.1 and 2).
This is why we make use of a logarithmic argument 
$F_L=\ln{(1+\delta F)}$ rather than of $\delta F$ itself.
That slow dependence prevents us from
reaching a close vicinity of the upper bound $d=4$, and thus
from clarifying an interesting question in Ref.\cite{1}:
if the dimension does asymptotically approach this bound or,
instead, saturates at some lower value of $d$
(for farther discussion see next Section).
The maximal value actually reached in our numerical experiments
was approximately $d\approx 3.73$ only due to a rapid increase
of the computation time as $d\to 4$.

The empirical dependence $d(F_L)$ (and hence $\Lambda_1(F_L)$
as well) has been constructed using the data for the two first groups
in Fig.1 with known $F_{cr}$ values. To this end, let us assume, first,
that $d(F_L)$ does reach the upper bound as $F_L\to\infty$.
Further, Fig.2 suggests that this dependence is close to linear
near $F_L=0$, hence we added the unity under logarithm in $F_L$.
Finally, we use a simple relation:
$$
   d(F_L)\,=\,2\,+\,\frac{4\,F_L}{F_0\,+\,2\,F_L}\,, \quad
   F_L=\ln{(1+\delta F)} \eqno (3.3)
$$
with a single fitting parameter $F_0$.

Surprisingly, the simple empirical relation (3.3) fits, with
a particular value of the fitting parameter $F_0=8.5$,
not only the points of the first two groups but also, with a
reasonable accuracy (especially for $d$), the rest of points 
with unknown $F_{cr}$ values set to zero in Fig.2.
Notice that deviations of empirical points from a smooth
dependence are not due to computation errors, which were carefully
checked, but apparently represent some fine structure of
chaotic attractors we did not study in the present work (cf. Ref.\cite{1}).

Still, the agreement between empirical data and analytical
relation is not yet satisfactory especially for the three
rightmost points with the biggest $F$.

For improving the agreement we used two methods.
First, we tried to fit the unknown $F_{cr}$ values for the three
groups of points around $F_L=7$ (see Fig.1 and the Table).
Interestingly, the two groups seem to belong to the same attractor
even though they are separated by a wide gap with limit cycles only.
Apparently, this is because of the fixed initial conditions (2.3) which
may or may not find themselves within the attractor basin for a particular
value of $F_L$ (see also Ref.\cite{2,1}). 
For this reason we cannot directly see the vicinity of chaos border
for these three groups of points. Hence, the two new values of $F_{cr}$
are the fitting parameters unlike the first two.

{\large
\begin{center}
\begin{tabular}{|c|c|c|c|c|}
\multicolumn{4}{l}{Table. Characteristics of few--freedom
chaotic attractors}\\ [3mm] \hline
\multicolumn{1}{|c|}{attractor's} &
\multicolumn{1}{|c|}{$F_{cr}$} &
\multicolumn{1}{|c|}{$F_{cr}/F$} &
\multicolumn{1}{|c|}{$R(F)$} &
\multicolumn{1}{|c|}{$S(F)$} \\ 
number & chaos border &  & size & shape \\ \hline
1 & $173$ & $1$ & $1.41$ & $0.74$ \\ \hline
2 & $190$ & $1$ & $1.43$ & $0.75$ \\
2 & $190$ & $0.32$ & $1.7$ & $0.9$ \\ \hline
3 & $1913$ & $0.83$ & $1.9$ & $1.0$ \\ \hline 
4 & $3038$ & $0.93$ & $2.0$ & $1.1$ \\ \hline  \hline
$F>4000$ & $0.8\,F$ & $0.8$ & $2 - 6.5$ & $1 - 1.6$ \\ \hline
\end{tabular}
\end{center}
}

Beyond the five groups we managed to compute just a few scattered points
which prevented the location of other attractors.
In this region we used another method based on a simple assumption
that each point belongs to a separate attractor whose chaos border
is directly related to the position of this point
$$
   F_{cr}(F)\,=\,F_s\cdot F \eqno (3.4)
$$
where $F_s$ is a new fitting parameter.
This method of a statistical nature is, in a sense, opposite 
to the first one. The assumption (3.4) is very crude, of course,
but it allows us to farther improve the agreement with
empirical relations (3.2) and (3.3) including two points
at $F_L\approx 13.5$ but not
the last one at the biggest $F_L\approx 14.5$.

Certainly, a couple of these 'abnormal' points is too few for any definite
conclusions. Yet, they may, and apparently do as we shall see below, indicate 
a different asymptotic behavior, as $F\to\infty$, compared to
the simple relations (3.2) and (3.3).
To clarify this important question we turned to the very limit
$F=\infty$. To reach this limit we may fix parameters
$f$ and $\mu$, and set $\gamma =\omega_2=0$ (see scaling (2.1)
and (2.2)).
In this way we arrive at a conservative system with 
the simple Hermitian Hamiltonian:
$$
   H_0(a_1,\,a_2)\,=\,\mu a_1^2a_2^*\,+\,fa_1^*\,
   +\,{\rm c.c.}\,=\,0 \eqno (3.5)
$$
where the latter equality is due to the initial conditions (2.3).

A few sample runs of this limit system revealed that all characteristics
(2.5) but the dimension $d$ of an energy surface now, which is the limit
of a chaotic attractor for a finite $\gamma$, kept growing with time.
This suggests an unbounded motion along some non--compact
energy surface, and moreover with the ever increasing dimensional 
Lyapunov exponent $\lambda_1\to\infty$ for $t\to\infty$.

As to the dimension of the energy surface itself it quickly
converges to the maximal value $d=4$.
At the first glance, it appears strange as one would expect this surface 
to be
three--dimensional. The explanation of the apparent paradox is
the following. The metric of the motion $M=(+,0,0,-)$ has now two zero
Lyapunov exponents: one corresponds to the eigenvector along a trajectoty,
as usual, while the other one is associated with the eigenvector 
across the energy
surface (the so--called marginal instability, see, e.g., Ref.\cite{6}). 
It means that this metric characterizes
an infinitely narrow four--dimentional layer around the energy surface
rather than the surface itself. Notice that the invariant measure 
of ergodic motion (the so--called microcanonical distribution) 
on energy surface
is determined by the phase--space volume in that layer, and not by area
on the surface.

Increasing $\lambda_1$ in the limit (3.5)
is at variance with the asymptotics of Eq.(3.2) for $\gamma\to 0$.
Indeed
$$
   \lambda_1\,=\,\Lambda_1\gamma\,\to\,\frac{|\Lambda_{\Gamma}|}{F_0}\cdot
   \gamma\ln{\left(\frac{F_s\mu f}{\gamma^2}\right)}\,\to 0\,, 
   \quad \gamma\to 0
   \eqno (3.6)
$$
To cope with this difficulty we need a power--law argument in 
the right--hand side of Eq.(3.3)
rather than a logarithmic one, $F_L$.
To this end, we simply added to the latter a power--law term in the form
$$
   F_L\,\to\,F_L\,+\,\exp{(\beta\,(F_L\,-\,F_1))} \eqno (3.7)
$$
with two parameters, $\beta$ and $F_1$. The latter 
characterizes the crossover to the asymptotic behavior where
the Lyapunov exponent is described by the relation
$$
   \lambda_1\,\to\,\frac{|\Lambda_{\Gamma}|}{F_0}\cdot{\rm e}^{-\beta F_1}\,
   \frac{(F_s\mu f)^{\beta}}{\gamma^{2\beta -1}}
   \eqno (3.8)
$$
The limit behavior in system (3.5) ($\lambda_1$ growth) implies
$\beta >1/2$.

There is an interesting possibility to calculate more accurate value of
$\beta$ using our preliminary empirical data for the limit (3.5).
The point is that conservative system (3.5) possesses only one quantity
of the frequency dimension, the inverse time. So, one may conjecture that 
the variable $1/t$
plays in conservative system (3.5) a role similar to that of $\gamma$ in
our main dissipative model (1.1). According to our preliminary data
all the motion characteristics
(2.5) but $d$ in the former case
are growing with time as a power law. Particularly, $\lambda_1\propto t^{1/3}$.
By comparison with Eq.(3.8) we obtain $\beta =2/3$
to be used below.

With all these improvements we show in Fig.3 our final fitting
the empirical data for the attractor dimension and Lyapunov exponents.

\begin{figure}[h]
\centerline{\epsfxsize=15cm \epsfbox{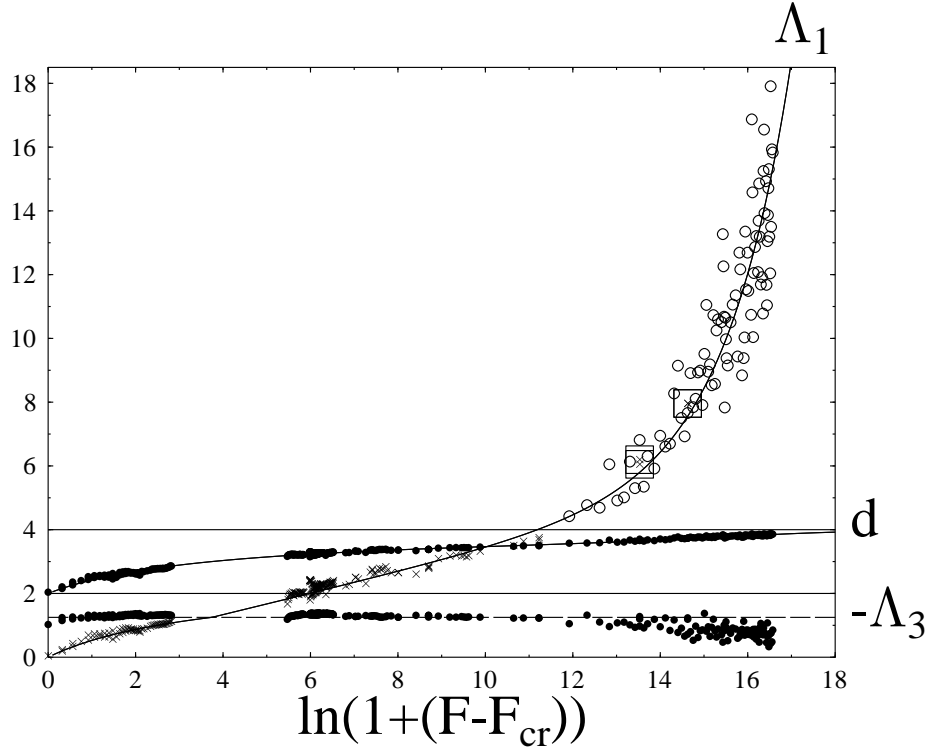}}
\caption{
         Same as in Fig.2 with two more identified attractors
         (see the Table), and the statistical estimate (3.4) for
         the rest.
         Two curves are old empirical relations (3.3) and (3.2) 
         with a new term (3.7) and fitting parameters: 
         $F_s=0.8,\ F_0=7.5,\ F_1=11.5$. Circles show new data
         for $\Lambda_1$ (see Section 4).
}
\end{figure}

The most interesting result of our studies is finding eventually the
two quite different chaotic structures with respect
to their dependence on the principal parameter $F$.
The most of our empirical data belong to the structure with the slow,
logarithmic, dependence.
It comprises the region 
from $\delta F=(\delta F)_1\sim 1$ up to the crossover
$$
   (\delta F)_2\,\approx\,{\rm e}^{F_1}\approx 10^5 \eqno (3.9)
$$
Beyond $(\delta F)_2$ the chaotic structure becomes 'turbulent'
with a relatively
fast, power--law, dependence on $F$.
The crossover between the two regions is clearly seen in Fig.3 
from the
$\Lambda_1(F_L)$ dependence, and can be noticed in $d(F_L)$ variation
as well. So far, we have no explanation as to a large value of this crossover.
However, notice that in the logarithmic scale $F_L$ the crossover
$(F_L)_2\approx F_1\sim 10$ is not that big.

Since the logarithmic dependence in Eq.(3.3) becomes linear for
small $\delta F$ there is actually the third region bounded from above
by one more crossover $(\delta F)_1\sim 1$. This region of threshold
chaos is most clearly seen in Fig.1.

All the regions can be descibed by unified and relatively simple
empirical relations (3.2), (3.3), and (3.7) to a reasonable accuracy.
An interesting peculiarity of these relations is the discontinuity of the
first derivative $s=d\Lambda_1(F_L)/dF_L$ at $d=3$.
The ratio of the slopes above to below this point
$$
   \frac{s^+}{s_-}\,=\,\frac{\Lambda_{\Gamma}}{\Lambda_3}\,\approx\,2
   \eqno (3.10)
$$
is independent of the function itself, similar to Eq.(3.2).

\begin{figure}[h]
\centerline{\epsfxsize=15cm \epsfbox{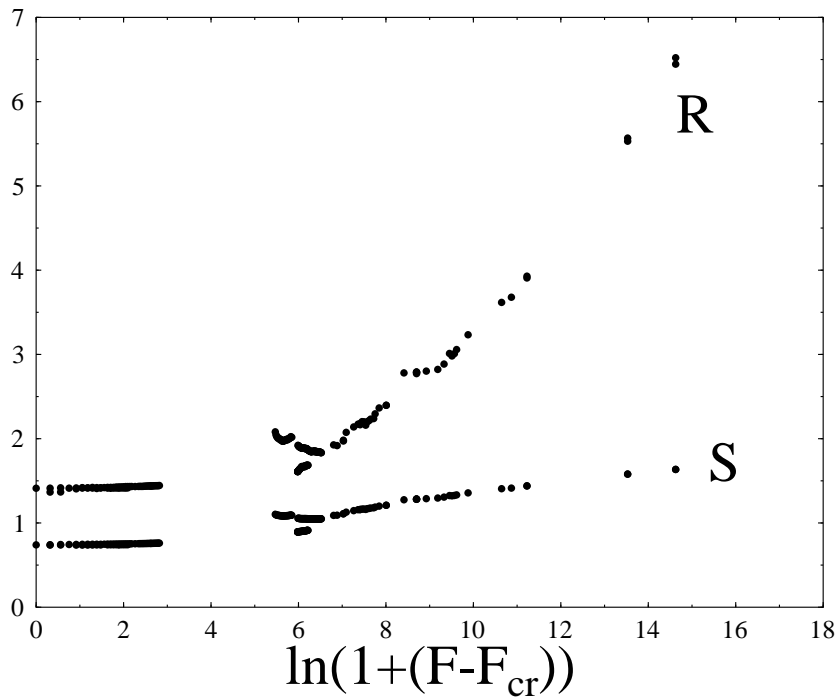}}
\caption{
         Same as in Fig.3 for the size $R$ (2.5c) (upper points)
         and shape $S$ (2.5d) (lower points).
}
\end{figure}

Finally, we display in Fig.4 the behavior of two geometrical characteristics
of chaotic attractor, the size $R$ (2.5c) and shape $S$ (2.5d).
We did not attempt to construct the empirical relations for these 
characteristics. However, a clear qualitative behavior of those is also
interesting and important. First, both $R$ and $S$ grow with $F_L$
which is also in agreement with the limit behavior in system (3.5).
In other words, the attractor becomes more and more large and elongated.
Another interesting point is in that the geometrical characteristics 
are multivalued that is specific, at the same value of $F_L$, for a particular
attractor.
This confirms that we, indeed, have found some different attractors
as summarized in the Table.

\section{Conclusion}
In the present work, which continues our previous studies \cite{1},
we investigate the structure of chaotic atractors in 
the Khalatnikov - Kroyter model of two freedoms (1.1). As was shown in 
Ref.\cite{2,1} this simple model describes the birth of the wave turbulence
(dynamical chaos) in a real very complicated physical phenomenon - the 
dynamics of the
second sound in helium II driven by an external force $f$.
The model is applicable for weak force only which proved to be sufficient
for producing chaos. It means that under particular conditions such a
few--freedom model can represent some features of the many--freedom 
turbulent structure. The main goal of the present studies was to find out
what would be peculiarities of the few--freedom 'turbulence' in a wide
range of the parameters, particularly of the driving force.
To put it other way, we were interested in the structure of 
the few--freedom chaos as compared to that of the multi--freedom one.

To this end, we have chosen a number of dimensionless characteristics
of the chaotic structure (2.5), und studied their dependence on a single
dimensionless parameter $F$ (2.1) in a series of numerical experiments.
We had expected some qualitative differences between 
the few-- and multi--freedom chaos, yet the surprise was in the nature
of the difference which proved to be the dependence
of chaos characteristics on the principal parameter $F$. 
While in the multi--freedom chaos this dependence is fast (a power law)
in a few--freedom chaos it is very slow (logarithmic). 
Even though our studies were restricted to a particular model we conjecture
that this principal difference is of a generic nature. Apparently, it is
related
to the strict limitation of the attractor dimension in a few--freedom system
(to the interval $2<d<4$ in our model).

In spite of satisfaction with this finding we went on in numerical experiments
trying to reach as large $F$ as possible in a reasonable computation time
(days), and it was not in vain! First, we tried to find some indication,
if any, of deviations from the original empirical relations which could be
interpreted as a lower asymptotic dimension ($d<4$) than assumed above ($d=4$).
Eventually, we have found the deviation but of the opposite sign
(see Fig.1 and 3)! Even though this is just three points we put forward
a different explanation - the transition to a new, power law, dependence
of the attractor characteristics which is similar to that in a 
multi--freedom system (Section 3).

Currently, we distinguish, altogether, the three regions
in the principal parameter $F$. In terms of $\Lambda_1$ they are as follows 
(see Figs.1 and 3, and Eqs.(3.2),(3.3),(3.7), and(3.9)):

(i) the threshold domain
$$
   \Lambda_1\,=\,-\Lambda_3\,(d\,-\,2)\,\approx\,\frac{4|\Lambda_3|}{F_0}
   \cdot (F-F_{cr})\,, \quad F_{cr}\,<\,F\,\lsim\,F_{cr}\,+\,1 \eqno (4.1a) 
$$
with a 'turbulent' behavior (cf. Ref.\cite{1});

(ii) the logarithmic region
$$
   F_{cr}\,+\,1\,\lsim\,F\,\lsim\,(\delta F)_2\,\approx 10^5 \eqno (4.1b)
$$
this is our main result revealing a qualitative difference between
the few-- and multi--freedom chaos, and

(iii) the asymptotics
$$
   \Lambda_1\,\approx\,\frac{|\Lambda_{\Gamma}|}{F_0}
   \cdot{\rm e}^{-2F_1/3}\cdot F^{2/3}\,, \quad
   F\,\approx\,\delta F\,\gsim\,(\delta F)_2 \eqno (4.1c)
$$
where the dynamics is similar again to a multi--freedom chaos ('turbulence', 
cf. the first
region (4.1a)) provided
unbounded motion in the conservative limit (3.5).

Remarkably, all the three regions can be described by a unified empirical
relation which is rather simple and reasonably accurate.
Particularly, this confirms the assumption that (rather than answer to
the question if) the dimension of chaotic attractor in the model considered
does reach the upper bound $d=4$ (see Eq.(3.3) and above).

At the time of publication Ref.\cite{7}, some
preliminary numerical experiments at very large 
$F\sim 10^7\ (f\approx 3\times 10^7)$ have presented the first evidence for
the transition to a power--law dependence in our few--freedom model. 
It was just three points for $\Lambda_1$ near the crossover (3.9) marked
in Fig.3 by squares.
By now, we have collected much more empirical data
at the crossover and beyond, up to $F\approx 7\times 10^7$. 
These are also shown in Fig.3 (circles), and perfectly fit, 
to a less accuracy though,
the previously found empirical dependence $\Lambda_1 (F-F_{cr})$. 
Each point in the new region costed about a full day of work for 
one processor on the ALPHA supercomputer.

Unlike this surprising agreement in extrapolation of $\Lambda_1$, 
the new data for $\Lambda_3$ show a systematic deviation down to
$<-\Lambda_3>\approx 0.75$ above crossover (3.9).
At the moment it is not clear if this decrease is caused by insufficient
computation accuracy, we had to permit in order to cope with the enormous
computation time, or a real transition for $\Lambda_3$ is observed, indeed.
In any case, the corresponding change in $\Lambda_1$ is negligible
in this region (see Eq.(3.2)).

There is an interesting possibility to compare the unexpected and unexplained
invariance of $\Lambda_3(F)$ in our model with the data in Ref.\cite{8}
for a close model of two freedoms also. 
In the latter work the fractal geometry of a chaotic attractor was studied
in detail. For the main set of model's parameters the Lyapunov exponents 
are
$$
   \lambda_n\,=\,3.34\,,\ 0\,,\ -1.79\,,\ -5.55
$$
with the same metric $M$ (2.4) as in our model, the dissipativity 
$\lambda_{\Gamma}=-4$, and the attractor dimension $d=3.28$ which is well
within the logarithmic region (see Fig.3).
Even more interesting is the ratio
$$
   \frac{\lambda_{\Gamma}}{\lambda_3}\,=\,2.23
$$
which differs from our result by 10\% only.

In the very conclusion, we would like to mention that an important question 
how univeral the described structure of the
few--freedom chaos might be, remains open and certainly deserves 
further studies. In particular, the above condition
for a power--law asymptotics (unbounded motion in conservative limit (3.5))
is hardly generic. Rather, it is characteristic for a more narrow class
of nonlinear models as compared to those with the wide logarithmic region.

\vspace{5mm}

{\bf Acknowledgments.} We are grateful to M. Kroyter and V.V. Parkhomchuk
for interesting discussions on, and stimulating interest to, this work.

\end{document}